\documentclass[conference, a4paper]{IEEEtran}
\IEEEoverridecommandlockouts
\usepackage{algorithm,algorithmic}
\usepackage{amsmath,amssymb,mathrsfs}
\usepackage{cite}
\usepackage{subfigure}
\usepackage{graphicx,color}
\usepackage[nomain,acronym,toc]{glossaries}%[nonumberlist,acronym,toc]
\usepackage{hyperref}
\usepackage{subfigure}
\usepackage{balance}
\usepackage{tikz,siunitx}
\usepackage{url}
\usepackage{graphicx} %use graph format
\usepackage{epstopdf}
\usepackage{lipsum} 

\newtheorem{defn}{Definition}
\newtheorem{lem}{Lemma}

\newtheorem{thm}{Theorem}
\newtheorem{cor}{Corollary}[thm]

\definecolor{sblue}{RGB}{0,51,120}
\definecolor{sred}{RGB}{200,51,130}

\renewcommand{\eqref}[1]{(\ref{#1})}
\newcommand{\figref}[1]{Fig. \ref{#1}}

\newacronym[description=User equipment]{ue}{UE}{user equipment}

\newacronym[description=Mobile edge computing]{mec}{MEC}{mobile edge computing}
\newacronym[description=Bit error rate]{ber}{BER}{bit error rate}
\newacronym[description=Correlation based incrementak task offloading]{cito}{C-ITO}{correlation based incrementak task offloading}
\newacronym[description=Incrementak task offloading]{ito}{ITO}{incrementak task offloading}
\newacronym[description=System level]{sl}{SL}{system level}
\newacronym[description=Robot level]{rl}{RL}{robot level}
\newacronym[description=Signal to noise ratio]{snr}{SNR}{signal to noise ratio}
\newacronym[description=Symbol error rate]{ser}{SER}{symbol error rate}
\newacronym[description=Symbiotic based robotic swarm system]{sbrss}{SBRSS}{symbiotic based robotic swarm system}
\newacronym[description=Network lifetime]{nl}{NL}{network lifetime}

\ifCLASSINFOpdf
\else
\fi

\begin{document}
	
	\title{\huge Robot Subset Selection for Swarm Lifetime Maximization in Computation Offloading with Correlated Data Sources\\}
	
	\author{Siqi Zhang, Na Yi, and Yi Ma$^\dagger$\\
		{\small 5GIC and 6GIC, Institute for Communication Systems, University of Surrey, Guildford, UK, GU2 7XH}\\
		{\small Emails: (s.zhang, n.yi, y.ma)@surrey.ac.uk}}
	\markboth{}%
	{}

	\maketitle
	
	\begin{abstract}
		Consider robot swarm wireless networks where mobile robots offload their computing tasks to a computing server located at the mobile edge.
		Our aim is to maximize the swarm lifetime through efficient exploitation of the correlation between distributed data sources.  
		The optimization problem is handled by selecting appropriate robot subsets to send their sensed data to the server. 
		In this work, the data correlation between distributed robot subsets is modelled as an undirected graph.
		A least-degree iterative partitioning (LDIP) algorithm is proposed to partition the graph into a set of subgraphs. 
		Each subgraph has at least one vertex (i.e., subset), termed representative vertex (R-Vertex), which shares edges with and only with all other vertices within the subgraph; only R-Vertices are selected for data transmissions. 
		When the number of subgraphs is maximized, the proposed subset selection approach is shown to be optimum in the AWGN channel.
		For independent fading channels, the max-min principle can be incorporated into the proposed approach to achieve the best performance. 
	\end{abstract}
	
	\section{Introduction}
	By robot swarm, we refer to the definition provided by the US Army Academy \cite[Sec. II]{8993024}:
	\begin{defn}[Robot Swarm]\label{def01}
		A robot swarm is a group of (three or more) robots that perform tasks cooperatively while receiving limited or no control from human operators.
		The term `cooperatively' is defined in the way that involves mutual assistance in working towards a common goal.
		It does not necessarily imply or require communication or explicit coordination between entities.
	\end{defn}
	This technology has found rich applications including disaster recovery, defense, reconnaissance, inspection, mapping, farming, food management, and space systems.
	
	In light of {\em Definition \ref{def01}}, we consider a robot swarm where a set of mobile robots cooperatively perform a single task. 
	Every robot can conduct environment sensing and if required send their sensed data to a computing server through a connected access point (AP).
	The server conducts processing on the received data and sends computing outcomes to the robots. 
	This work falls into the scope of mobile edge computing (MEC) in the wireless communication domain.
	The motivation of offloading computing tasks to the edge server lies in (see \cite{9139976}): 
	{\it 1)} data processing (computing) tasks are often energy hungry and thus not suitable for battery-powered mobile robots;
	{\it 2)} edge servers are often equipped with powerful computing units (such as GPUs, APUs or NPUs), which can provide parallel and low-latency computing services. 
	
	Our prior arts analysis shows that most of current MEC computation-task offloading works are focused on resource allocation problems in single-user equivalent computing scenarios (e.g. \cite{9308935,7938331,9448864,8955967}), 
	i.e., despite multi-user communications in reliability, computing tasks (or data) from different devices are processed individually at the edge server. Recently, multi-user edge computing has attracted increasing interests, where data from different devices must be jointly processed to yield the computing outcome. 
	In this case, users are branch and bound in the procedure of computation-task offloading \cite{9815185}. 
	
	This work falls into the scope of multi-user edge computing, with the aim of maximizing the lifetime of robot swarm exploiting the data source correlation between mobile robots. 
	Major contributions of this work include: 
	{\em 1)} the development of a novel robot-subset model using an undirected graph, describing the data source correlation between spatially distributed robots;
	{\em 2)} the establishment of a novel cost function, formulating the optimization problem for the swarm-lifetime maximization;
	{\em 3)} the development of a novel graph partitioning and subset selection approach to tackle the swarm-lifetime maximization problem. 
	It is shown that, when the number of subgraphs is maximized, the proposed subset selection approach is optimum in the AWGN channel.
	As far as independent flat-fading channels are concerned, the max-min principle can be incorporated into the proposed approach to maximize the swarm lifetime. 
	The above findings are elaborated through well-designed computer simulations.
	
	\section{System Model and Problem Formulation}
	In this section, we introduce the basic concept of robot subset, data correlation, communication model, as well as the formulation of optimization problem.
	
	\subsection{Robot, Subset, and Data Correlation Model}
	Consider a robot swarm where a set of mobile robots ($N$) collaboratively perform a single task. 
	Mathematically, a robot is defined as a tuple, $s\triangleq(x, \mathcal{E}, \varepsilon)$, 
	where $x$ stands for the sensed data, $\mathcal{E}$ for the remaining energy of the robot, and $\varepsilon$ for the energy consumption per job \footnote{For the sake of concise presentation, $\varepsilon$ is assumed to be a constant.}. 
	Then, the robot swarm is described as a set: $\mathcal{S}\triangleq\{s_n|n=1, ...,N\}$.
	Provided that all robots send their data to the MEC server, the multi-user edge computing process can be described as
	\begin{equation}\label{eqn01}
		y=f(x_n|n=1, ..., N),~ y\in\mathcal{A}\triangleq\{a_l|l=1, ..., L\},
	\end{equation}
	where $x_n$, $_{n=1,...,N},$ denotes the data from the $n^{th}$ robot, and $y$ denotes the computing outcome.
	Following {\em Definition \ref{def01}}, we assume there is no human intervention in the computing and decision making.
	Then, the outcome $(y)$ must belong to a finite-alphabet set defined by $\mathcal{A}$,
	which specifies all possible actions (i.e., $a_l, _{l=1,\cdots,L}$) of the robot swarm; $L$ is the total number of actions. 
	Since $(\mathcal{E}, \varepsilon)$ has no impact on $y$, \eqref{eqn01} can have the following notation-simplified form 
	\begin{equation}\label{eqn02}
		y=f(\mathcal{S}),~ y\in\mathcal{A}.
	\end{equation}
	
	The above description is for the generic case.
	However, in many practical applications, the MEC does not need the whole set ($\mathcal{S}$) to form the computing outcome.
	This is because robots' data are correlated and carry considerable redundancy. 
	This insight motivates us to propose the concept of robot subset in the multi-user edge computing. 
	
	\begin{defn}[Robot Subset]\label{def2}
		Consider a subset of robots within the swarm, denoted by $\Omega_m\subset\mathcal{S}$ with $|\Omega_m|=K_m$, $_{m=1,..., M}$, where $|\cdot|$ stands for the cardinality of the set, and $M$ for the number of subsets. 
		The MEC outcome corresponding to the subset $\Omega_m$ is described by
		\begin{equation}\label{eqn03}
			y_m=f(\Omega_m),~ y_m\in\mathcal{A}.
		\end{equation}
		It is assumed that there exists a $K(<N)$, for $K_m\geq K$, the result $(y_m=y)$ always holds \footnote{This assumption turns out to be true for robot (or sensor) networks where robots (or sensors) are closely located in a geographical area (see \cite{8786812}).}.
		Using the terminology of data science \cite{7295461}, the robot subset ($\Omega_m$) is said to have the same value of information as the whole set  $(\mathcal{S})$. 
	\end{defn}
	
	{\em Definition \ref{def2}} implies that the whole set ($\mathcal{S}$) can be divided into $M$ robot subsets, with each having the same value of information. 
	Then, for each computing task, only one of subsets is required to send their sensed data to the MEC.
	This will result in significant saving in communication energy and longer swarm lifetime. 
	The issue of how to divide $\mathcal{S}$ into subsets will be addressed in Section \ref{sec4a}.
	
	\subsection{Communication Model}\label{sec2b}
	For every single task, the MEC chooses one of robot subsets (denoted by $\Omega_m$), requesting them send their data to the MEC through $|\Omega_m|$ orthogonal sub-channels.
	Without loss of generality, we assume $|\Omega_m|=K, {\forall m}$. 
	For a robot $s_k\in\Omega_m$, the channel capacity between this robot and the AP is denoted as
	$C(\alpha_k, p_k)$, where $\alpha_k$ is the channel quality, and $p_k$ is the transmit power which is capped by a physical constraint $(p_k\leq P_k)$. 
	Then, the robot subset ($\Omega_m$) must fulfill the following criteria:
	
	{\em Criterion 1:} $\forall s_k\in\Omega_m$, their transmission rate (denoted by $R_k$) must fulfill
	\begin{equation}\label{eqn04}
		R_k<C(\alpha_k, p_k)\leq C(\alpha_k, P_k);
	\end{equation}
	or otherwise the robot ($s_k$) or the subset ($\Omega_m$) must not be chosen due to the capacity outage. 
	
	{\em Criterion 2:}
	The data transmission must be completed within a time duration ($T_\textsc{c}$) for the sake of latency. 
	
	{\em Criterion 3:}
	All robots must have enough remaining energy to execute the allocated job. 
	
	After the computing process \eqref{eqn03}, the outcome ($y$) is broadcasted to all robots. 
	It is assumed that the broadcast is reliable and efficient. 
	This is the commonly used assumption in the literature (e.g., \cite{9308935,9760072,8638800}), which allows us focusing on the key problem of interest.
	
	\subsection{Optimization Problem}
	This work aims to maximize the swarm lifetime through optimum selection of robot subsets for the data transmission.
	The swarm lifetime is defined as:
	\begin{defn}[Swarm Lifetime]\label{def3}
		A task assigned to the robot swarm requires collaborative work amongst all the robots. 
		Should any robot stop working due to the short of device energy results in task failure of the whole robot swarm. 
		Therefore, the swarm lifetime counts as the maximum number of tasks a robot swarm can complete all the way from the beginning to the end. 
	\end{defn}
	
	Consider the life cycle of the $i^{th}$ task conducted by the robot swarm.
	Robots have the following energy consumption model:
	\begin{subequations}\label{eqn05}
		\begin{equation}\label{eqn05a}
			\mathcal{E}_k(i)=\mathcal{E}_k(i-1)-p_k(i)T_\textsc{c}-\epsilon-\varepsilon,~ \forall s_k\in\Omega(i),
		\end{equation}
		\begin{equation}\label{eqn05b}
			\mathcal{E}_\ell(i)=\mathcal{E}_\ell(i-1)-\varepsilon,~ \forall s_\ell\in\overline{\Omega(i)},
		\end{equation}
	\end{subequations}
	where $\epsilon$ is the energy consumption for communication-related computation such as the source and channel coding, 
	and the subscript $(\cdot)_\ell$ is the index for unselected robots.
	Note that we shall have the condition: 
	\begin{equation}\label{eqn06}
		\mathcal{E}_k(i)\geq0,  \mathcal{E}_\ell(i)\geq 0, \forall k, \ell;
	\end{equation} 
	or otherwise, the $i^{th}$ task won't happen, and the $(i-1)^{th}$ task would be the last task assigned to the robot swarm. 
	Therefore, whether or not we will have the $i^{th}$ task, this depends on the strategy of robot selection in the previous tasks. 
	
	Define $\Phi(i-1)\triangleq(\Omega(1), \Omega(2), ..., \Omega(i-1))$ with $\Phi(0)=\emptyset$.
	We further define $\Gamma(\Phi(i-1))$ the strategy of robot selection which yields
	\begin{equation}\label{eqn07}
		\Gamma(\Phi(i-1))=\left\{
		\begin{array}{cc}
			i ,& \eqref{eqn06}~\mathrm{is~true}\\
			0, & \eqref{eqn06}~\mathrm{is~false}
		\end{array}
		\right.
	\end{equation}
	
	According to {\em Definition \ref{def3}}, the swarm-lifetime maximization problem has the following objective function
	\begin{equation}\label{eqn08}
		\Phi^\star(i-1)=\underset{\Phi(i-1)}{\arg\max}~\Gamma(\Phi(i-1)).
	\end{equation}
	This is the optimization problem we strive to address in Section \ref{sec03}. 
	
	In the scope of wireless sensor networks (WSNs), the problem of system lifetime maximization has already been extensively studied; see the survey paper \cite{7812629} and the references therein. 
	However, the lifetime of WSN is very different from that of robot swarm in manifold: {\em 1)} WSNs have no task to execute, i.e., the term $\varepsilon$ won't appear in the energy cost function \eqref{eqn05}; {\em 2)} the life of WSN continues when one of sensors dies, i.e., the condition \eqref{eqn06} does not necessarily hold; {\em 3)} sensors often have their data highly or fully correlated, and thus the lifetime maximization problem can be handled through a sensor-level selection protocol; 
	there is no multi-user edge computing problem involved. 
	All of these differences render our robot swarm lifetime maximization a novel problem of optimization. 
	
	\section{The Method for Robot Swarm Lifetime Maximization}\label{sec03}
	\subsection{Analysis of The Optimization Problem}
	The implication of \eqref{eqn08} is that the lifetime maximization is equivalent to the convergence problem of an iterative algorithm.  
	The only difference is that we want the convergence to be as slow as possible.
	Specifically, \eqref{eqn05b} can be written into
	\begin{equation}\label{eqn09}
		\mathcal{E}_\ell(i)=\mathcal{E}_\ell(0)-i\varepsilon,~ \forall s_\ell\in\overline{\Omega}.
	\end{equation}
	It means that $\mathcal{E}_\ell(i)$ features linear convergence with the rate of $\varepsilon$. 
	A bit complicated case is \eqref{eqn05a}, where $E_k(i)$ does not converge linearly. 
	Nevertheless, for every single robot, we can present their energy consumption in an universal model
	\begin{equation}\label{eqn10}
		\mathcal{E}_n(i)=\mathcal{E}_n(0)-\zeta_n(i)-i\varepsilon,
	\end{equation}
	and $\zeta_n(i)$ is defined by
	\begin{equation}\label{eqn11}
		\zeta_n(i)\triangleq\sum_{i'=0}^{i}\sum_{j\in\Psi_n}(p_k(i')T_\textsc{c}+\epsilon)\delta(i'-j),
	\end{equation}
	where $\Psi_n$ is the set collecting all iteration indexes when the $n^{th}$ robot is chosen to send their data, 
	and $\delta(\cdot)$ is the Dirac Delta function \cite{proakis2001digital}.
	
	Define $e_n(i)\triangleq \mathcal{E}_n(0)-\zeta_n(i)$.
	A robot with smaller $e_n(i)$ has a shorter lifetime. 
	According to {\em Definition \ref{def3}}, the swarm lifetime is determined by the robot with the shortest lifetime.
	Therefore, the problem of swarm lifetime maximization is equivalent to maximizing the minimum of $e_n(i), \forall n$, i.e.,
	\begin{equation}\label{eqn12}
		\Phi^\star(i-1)=\underset{\Phi(i-1)}{\arg\max}\Big(\min (e_1(i), e_2(i), ..., e_N(i))\Big).
	\end{equation}
	The max-min optimization problem in \eqref{eqn12} is well-known as NP-hard (see \cite{6786314}). 
	However, it is possible to find good sub-optimum solutions based on some relaxed conditions. 
	We will tackle this problem step by step all the way from the AWGN channel to the fading channel.
	
	\subsection{Optimization in The AWGN Channel}
	In the context of AWGN channel, the channel quality measure, $\alpha_n$, is not robot dependent. 
	Instead, it can be simply set as: $\alpha_n=1, \forall n$.
	Then, the channel capacity is not a criterion for selecting a robot subset.
	The key parameters are mainly the remaining energy $\mathcal{E}_n$ and the subsets $\Omega_m, \forall n, m$. 
	Next, we will discuss about two cases when robots have or do not have identical initial state of the remaining energy.  
	
	\subsubsection{The case with identical state of the remaining energy}
	\begin{lem}\label{lemma01}
		Suppose: {c1)} $\mathcal{E}_n(0), \forall n,$ are identical; 
		{c2)} $\Omega_c\triangleq\Omega_1\cap\Omega_2\cdots\cap\Omega_M\neq\emptyset$.
		The swarm lifetime ($i$) is given by
		\begin{equation}\label{eqn13}
			i=\left\lfloor\frac{\mathcal{E}(0)}{pT_\textsc{c}+\epsilon+\varepsilon}\right\rfloor,
		\end{equation}
		where $\mathcal{E}$ and $p$ have their index omitted since they are identical for all robots, 
		and $\lfloor\cdot\rfloor$ stands for the integer floor.
	\end{lem}
	\begin{IEEEproof}
		Given the condition c1), robot subsets are the only influencing factors for the lifetime maximization.
		Given the condition c2), no matter which subset is chosen, the common part of all subsets ($\Omega_c$) is always chosen. 
		Therefore, robots within $\Omega_c$ have the shortest lifetime, and their energy consumption model \eqref{eqn10} becomes
		\begin{equation}\label{eqn14}
			\mathcal{E}(i)=\mathcal{E}(0)-i(pT_\textsc{c}+\epsilon+\varepsilon).
		\end{equation}
		Given $\mathcal{E}(i)\geq 0$, solving this inequality leads to \eqref{eqn13}.
	\end{IEEEproof}
	\begin{lem}\label{lemma02}
		Given the conditions c1) and c3) $\Omega_{m_1}\cap\Omega_{m_2}=\emptyset, \forall m_1\neq m_2$, 
		the swarm lifetime ($i$) is upper-bounded by
			\begin{equation}\label{eqn15}
				i\leq\frac{\mathcal{E}(0)-\left\lceil(i)/(M)\right\rceil(pT_\textsc{c}+\epsilon)}{\varepsilon},
			\end{equation}
	\end{lem}
	where $\lceil\cdot\rceil$ stands for the integer ceiling. 
	\begin{IEEEproof}
		With the conditions c1) and c3), the best strategy for subset selection is round robin because all subsets are equal. 
		In this case, the energy consumption model generally stays the same as \eqref{eqn10}, but the term $\zeta_n(i)$ becomes a robot-index independent term
		\begin{equation}\label{eqn16}
			\zeta(i)=\left\lceil\frac{i}{M}\right\rceil(pT_\textsc{c}+\epsilon).
		\end{equation}
		This expression describes the subsets which have been chosen for the most of times. 
		In other words, robots within those subsets have the shortest lifetime.
		Due to $\mathcal{E}(i)\geq 0$, we immediately have the inequality \eqref{eqn15}.
		The upper bound \eqref{eqn15} does not offer a closed-form solution for $i$.
		Nevertheless, it is very easy to determine $i$ through line searching. 
	\end{IEEEproof}
	
	{\em Lemma \ref{lemma01} \& \ref{lemma02}} exhibits the maximum swarm lifetime only for two special cases of robot subsets.
	Nevertheless, they lay the foundation for us to address the lifetime maximization in a generic case, for which we model the subset selection as a graph partitioning problem. 
	
	\begin{defn}[Graphical Model of Subsets]\label{def4}
		Define an undirected graph $G(V, E)$, where $V$ denotes the set of vertices and $E$ the set of edges. 
		There are $M$ vertices in the graph (i.e., $|V|=M$) which are corresponding to $M$ robot subsets. 
		The edge between two vertices reflects the presence of common part between two corresponding subsets (i.e., 
		the edge between the vertices $m_1$ and $m_2$ exists when $\Omega_{m_1}\cap\Omega_{m_2}\neq\emptyset$).
		A degree matrix $\mathbf{D}$ is formed to record the number of edges linked to vertices (see the definition of $\mathbf{D}$ in the graph theory \cite{west2001introduction}). 
	\end{defn}
	
	\begin{defn}[Graph Partitioning Problem]\label{def5}
		Our graph partitioning problem is stated by: partitioning an undirected graph into a maximal number of subgraphs ($\overline{M}$) fulfilling the criterion: c4) each subgraph has at least one vertex that shares edges with and only with all other vertices within the subgraph. 
		We call this vertex {\em representative vertex (R-Vertex)}.
	\end{defn}
	
	According to {\em Definition \ref{def4}}, the condition c2) can now be described by a complete graph $G(M, \frac{(M)(M-1)}{2})$, and the condition c3) can be described by a null graph $G(M, 0)$. 
	For the complete graph, we cannot further partition the graph, and thus have $\overline{M}=1$. 
	For the null graph, it is naturally partitioned into $\overline{M}=M$ subgraphs.
	For both cases, any vertex is a R-Vertex.
	When conducting subset selection at the subgraph level, it is trivial to find that the swarm lifetime is determined by \eqref{eqn14} and \eqref{eqn15}, respectively. 
	Applying the graph partitioning concept onto the generic case, the following result can be obtained. 
	
	\begin{thm}\label{thm01}
		For the generic case when the conditions c1) and c4) hold, the swarm lifetime is upper-bounded by 
		\begin{equation}\label{eqn17}
			i\leq\frac{\mathcal{E}(0)-\left\lceil(i)/(\overline{M})\right\rceil(pT_\textsc{c}+\epsilon)}{\varepsilon}.
		\end{equation}
	\end{thm}
	\begin{IEEEproof}
		Suppose that a graph can be maximally partitioned into $\overline{M}$ subgraphs following the criterion specified in c4). 
		Subset selection is applied on the subgraph level in a round robin manner. 
		For each subgraph, only (one of) the R-Vertex (R-Vertices) is chosen for the data transmission. 
		Since R-Vertices from different subgraphs are not directly connected to each other, 
		this is the equivalent case to that in {\em Lemma \ref{lemma02}}.
		Following the discussion in the proof of {\em Lemma \ref{lemma02}}, the swarm lifetime is determined by \eqref{eqn17}.
		
		The reason of choosing R-Vertices is to ensure that there is no overlap between selected subsets; 
		or otherwise, some robots would be chosen multiple times in one round, and such will shorten the swarm lifetime; as have already been discussed in {\em Lemma \ref{lemma02}}.
		Given $\overline{M}$ being the maximum number of subgraphs, the right-hand side of \eqref{eqn17} is maximized.
		{\em Theorem \ref{thm01}} is therefore proved. 
	\end{IEEEproof}
	
	{\em Theorem \ref{thm01}} shows that the key for lifetime maximization is to find $\overline{M}$ through an optimum graph-partitioning algorithm. 
	However, graph partitioning is usually a NP-hard problem, and we can hardly claim an optimum solution \cite{wu2019approximation}.
	In this paper, we propose a least-degree iterative partitioning (LDIP) algorithm that can offer a good sub-optimum solution for the graph partitioning. 
	
\begin{figure}[t]
		\centering
		\includegraphics[scale=0.28]{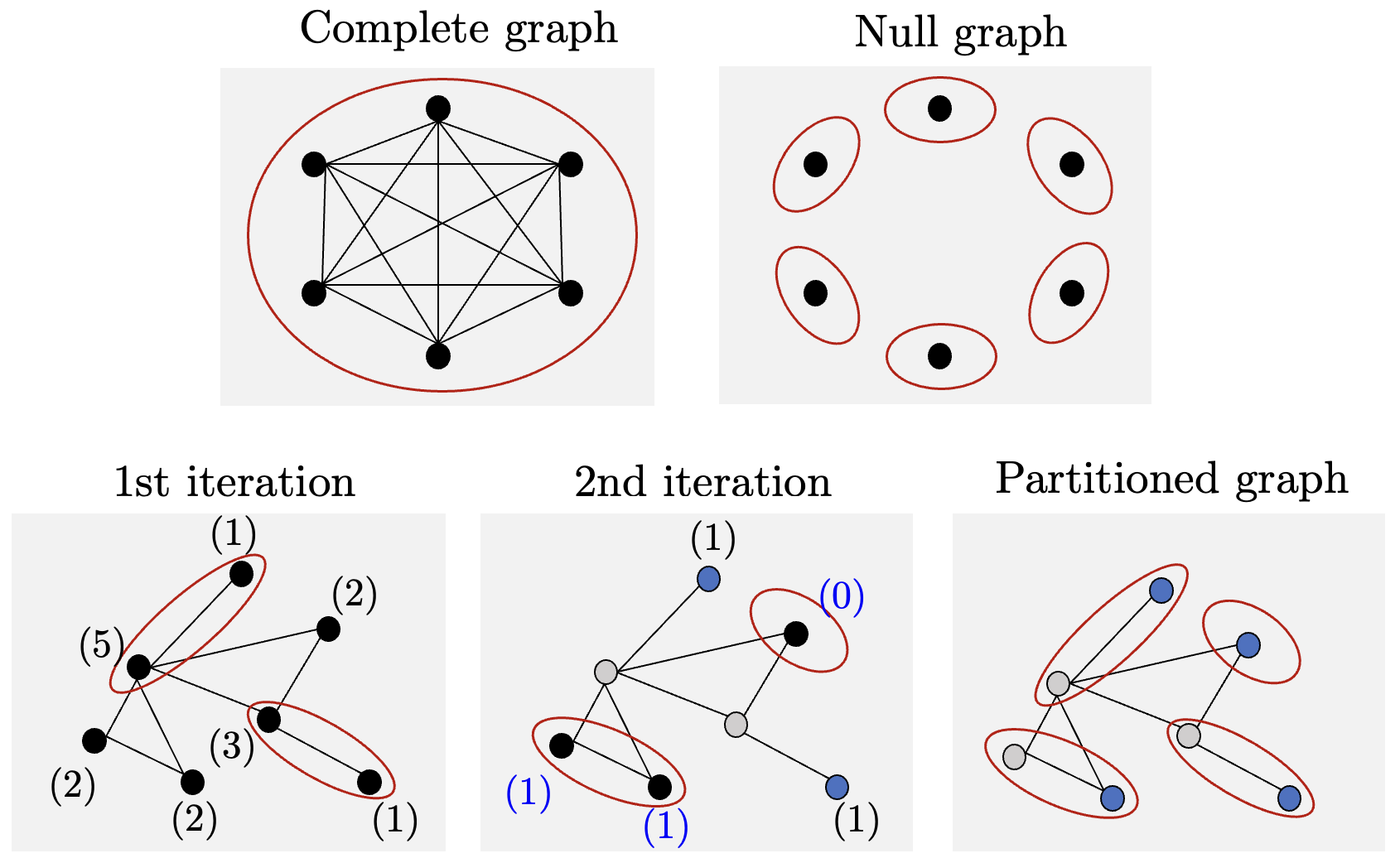}
		\caption{The concept of the LDIP algorithm for graph partitioning. 
			Red ellipses are subgraphs; blue-faced circles are R-Vertices of their subgraphs. }
		\label{fig2}
		\vspace{-1.5em}
	\end{figure}

	Fig. \ref{fig2} illustrates the concept of the LDIP algorithm. 
	Basically, the algorithm consists of three steps:
	
	{\bf\em Step 1 (Least degree finding):} Sort the diagonal of the degree matrix $\mathbf{D}$ in an ascent order; the first diagonal entry of $\mathbf{D}$ (denoted by $D_{(0,0)}$) has the least degree.
	
	{\bf\em Step 2 (Subgraph forming):} The vertex (subset) corresponding to $D_{(0,0)}$ is chosen to be the R-Vertex of the subgraph under construction. All vertices that are directly connected to the R-Vertex are included into this subgraph. 
	
	{\bf\em Step 3 (Degree matrix updating):} Update the degree matrix $\mathbf{D}$ (dimensional reduction) by eliminating all vertices that have been chosen at {\em Step 2}. Repeat {\em Step 1} until $\mathbf{D}$ reduces to a scalar equaling to 0, i.e., there is no vertex left. 
	
	The sub-optimality of LDIP lies in the fact that there may exist multiple vertices sharing the least degree at each iteration. 
	An optimum solution requires to visit all possible cases for each iteration, resulting in exponential searching complexity. 
	Instead, LDIP randomly chooses a vertex at each iteration, offering linear searching complexity at the price of optimality. 
	
	\subsubsection{The case with non-identical state of the remaining energy}
	Consider the condition: c5) $\mathcal{E}_n(0), \forall n$, can be different, the lifetime maximization problem becomes much more complicated because it involves multiple parameters (i.e., $\mathcal{E}_n(0)$ and $\Omega_m$) in the optimization.
	The round-robin algorithm is no longer optimum since subsets should not be equally treated. 
	Again, we start our study from the two special cases described in {\em Lemma \ref{lemma01} \& \ref{lemma02}}.
	
	\begin{thm}\label{thm02}
		Provided the conditions c2) and c5), the swarm lifetime ($i$) is upper-bounded by
		\begin{equation}\label{eqn18}
			i\leq\left\lfloor\frac{\min(\mathcal{E}_{\Omega_c}(0))}{pT_\textsc{c}+\epsilon+\varepsilon}\right\rfloor,
		\end{equation}
		where $\min(\mathcal{E}_{\Omega_c}(0))$ is corresponding to the robot in the set $\Omega_c$ who has the minimum $\mathcal{E}(0)$.
		A necessary condition for this upper bound to be achievable is 
		\begin{equation}\label{eqn19}
			\left\lfloor\frac{\min(\mathcal{E}_{\Omega_c}(0))}{pT_\textsc{c}+\epsilon+\varepsilon}\right\rfloor\leq
			\sum_{m=1}^M\left\lfloor\frac{\min(\mathcal{E}_{\Omega_m\setminus\Omega_c}(0))-i\varepsilon}{pT_\textsc{c}+\epsilon}\right\rfloor.
		\end{equation}
	\end{thm}
	\begin{IEEEproof}
		We start by proving the upper bound \eqref{eqn18}. 
		Following the proof of {\em Lemma \ref{lemma01}}, no matter which robot subset is chosen, 
		robots within the set $\Omega_c$ will always be selected, and they have the energy consumption model like \eqref{eqn14}, i.e.,
		\begin{equation}\label{eqn20}
			\mathcal{E}_\ell(i)=\mathcal{E}_\ell(0)-i(pT_\textsc{c}+\epsilon+\varepsilon), \ell\in\Omega_c.
		\end{equation}
		Then, the lifetime of $\Omega_c$ is limited by the robot with the minimum $\mathcal{E}(0)$, which has the form \eqref{eqn18}.
		According to {\em Definition \ref{def3}}, the swarm lifetime cannot go beyond this upper limit. 
		
		We further consider robots in the set $\Omega_m\setminus\Omega_c, \forall m$.
		Their energy consumption model is given by
		\begin{equation}\label{eqn21}
			\mathcal{E}_k(i)=\mathcal{E}_k(0)-w_m(pT_\textsc{c}+\epsilon)-i\varepsilon,~ k\in\Omega_m\setminus\Omega_c,
		\end{equation}
		where $w_m$ denotes how many times the robot subset is chosen. 
		Then, the lifetime of the robot set $\Omega_m\setminus\Omega_c$ is given by
		\begin{equation}\label{eqn22}
			w_m=\left\lfloor\frac{\min(\mathcal{E}_{\Omega_m\setminus\Omega_c}(0))-i\varepsilon}{pT_\textsc{c}+\epsilon}\right\rfloor.
		\end{equation}
		Consider an optimistic case when \footnote{When \eqref{eqn23} does not hold, the common part of two sets will be called whenever one of them are called. This can reduce the lifetime of the union of the two sets.}
		\begin{equation}\label{eqn23}
			(\Omega_{m_1}\setminus\Omega_c)\cap(\Omega_{m_2}\setminus\Omega_c)=\emptyset, \forall m_1\neq m_2,
		\end{equation}
		the lifetime for the union set $(\Omega_1\setminus\Omega_c)\cup(\Omega_2\setminus\Omega_c)\cup\cdots(\Omega_M\setminus\Omega_c)$ is $\sum_{m=1}^{M} w_m$. 
		This lifetime must be no less than the upper bound \eqref{eqn18}; or otherwise, 
		the upper bound is not achievable. {\em Theorem \ref{thm02}} is therefore proved.
	\end{IEEEproof}
	
	Note that the proof of {\em Theorem \ref{thm02}} has already incorporated the condition c3), which is equivalent to \eqref{eqn23} with $\Omega_c=\emptyset$. 
	Hence, for the condition c3), we have
	\begin{equation}\label{eqn24}
		w_m=\left\lfloor\frac{\min(\mathcal{E}_{\Omega_m}(0))-i\varepsilon}{pT_\textsc{c}+\epsilon}\right\rfloor
	\end{equation}
	and reaches the following conclusion:
	\begin{cor}\label{cor01}
		Given the conditions c3) and c5), the swarm lifetime ($i$) is upper-bounded by
		\begin{equation}\label{eqn25}
			i\leq
			\sum_{m=1}^M\left\lfloor\frac{\min(\mathcal{E}_{\Omega_m}(0))-i\varepsilon}{pT_\textsc{c}+\epsilon}\right\rfloor.
		\end{equation}
	\end{cor}
	
	The technical insight from {\em Theorem \ref{thm02}} and {\em Corollary \ref{cor01}} is that robot subsets must be weighted in the procedure of subset selection. 
	Their weights can be quantified by $w_m$ (see \eqref{eqn22} and \eqref{eqn24}), which are related to the terms $\min(\mathcal{E}_{\Omega_m}(0))$ and/or $\min(\mathcal{E}_{\Omega_m\setminus\Omega_c}(0))$.
	This insight is also applicable to the generic case, where R-Vertices must be weighted.  
	After the graph partitioning, we denote $\Omega_{\overline{m}}$ to be the robot subset corresponding to the R-Vertex of the $\overline{m}^{th}$ subgraph. 
	Since R-Vertices are disconnected from each other, we can follow {\em Corollary \ref{cor01}} to determine the swarm lifetime as
	\begin{equation}\label{eqn26}
		i\leq
		\sum_{\overline{m}=1}^{\overline{M}}\omega_{\overline{m}},~ \omega_{\overline{m}}=\left\lfloor\frac{\min(\mathcal{E}_{\Omega_{\overline{m}}}(0))-i\varepsilon}{pT_\textsc{c}+\epsilon}\right\rfloor.
	\end{equation}
	In addition, \eqref{eqn26} tells us, when the graph partitioning yields multiple results with the same $\overline{M}$, the one maximizing the upper bound \eqref{eqn26} should be chosen for the sake of lifetime maximization.
	
	\subsection{Optimization in Independent Flat-Fading Channels}
	Consider the channel quality $(\alpha_n)$ to be a random variable that is independent but not necessarily identical with respect to the robot index $(n)$ and the task index $(i')$.
	Then, the transmit power ($p$) also varies independently. 
	In this case, the energy consumption model for the $n^{th}$ robot goes back to the original version \eqref{eqn10}, 
	and the term $\zeta_n(i)$ can be written into 
		\begin{IEEEeqnarray}{ll}\label{eqn27}
			\zeta_n(i)&=T_\textsc{c}\sum_{i'=0}^{i}\sum_{j\in\Psi_n}(p_n(i'))+|\Psi_n|\epsilon,\\
			&=w_nT_c\underbrace{\left(\frac{1}{w_n}\sum_{i'=0}^{i}\sum_{j\in\Psi_n}(p_n(i'))\right)}_{\triangleq \overline{p}_n}+w_n\epsilon.
	\end{IEEEeqnarray}
	Following the discussion leading to {\em Theorem \ref{thm02}} and {\em Corollary \ref{cor01}}, it is straightforward to obtain 
		\begin{equation}\label{eqn29}
			w_n=\left\lfloor\frac{\mathcal{E}_{n}(0)-i\varepsilon}{\overline{p}_nT_\textsc{c}+\epsilon}\right\rfloor.
	\end{equation}
	
	In general, the averaged power $\overline{p}_n$ is robot dependent. 
	The lifetime of R-Vertices is limited by the robot with the smallest $w_m$.
	Hence, the swarm lifetime is given by
	\begin{equation}\label{eqn30}
		i\leq
		\sum_{\overline{m}=1}^{\overline{M}}\omega_{\overline{m}},~ \omega_{\overline{m}}=\min\left\{\left\lfloor\frac{\mathcal{E}_{n}(0)-i\varepsilon}{\overline{p}_nT_\textsc{c}+\epsilon}\right\rfloor, n\in\Omega_{\overline{m}}\right\}.
	\end{equation}
	Then, the lifetime maximization is equivalent to maximizing the upper bound of \eqref{eqn30}, i.e., following the max-min principle as stated in \eqref{eqn12}
	\begin{IEEEeqnarray}{ll}
		i&\leq\max\left(\sum_{\overline{m}=1}^{\overline{M}}\omega_{\overline{m}}\right)\\
		&\leq\sum_{\overline{m}=1}^{\overline{M}}\max_{p_n(i')}\left(\min\left\{\left\lfloor\frac{\mathcal{E}_{n}(0)-i\varepsilon}{\overline{p}_nT_\textsc{c}+\epsilon}\right\rfloor, n\in\Omega_{\overline{m}}\right\}\right).\label{eqn32}
	\end{IEEEeqnarray}
	Again, we emphasize that the max-min problem in \eqref{eqn32} is NP-hard. 
	Nevertheless, we can employ the max-min algorithm proposed in \cite{8955967} to handle the optimization problem in \eqref{eqn32},
	while the following criteria must be followed in the subset selection:
	
	{\em 1) Criterion 1} in Section \ref{sec2b} specifies the cap of transmit power. 
	Any subset that does not fulfill {\em Criterion 1} is not chosen for data transmission. 
	
	{\em 2)} When none of R-Vertices satisfy {\em Criterion 1}, we choose a subset from those which are not R-Vertices. 
	
	By this means, we incorporate the max-min principle into the LDIP algorithm to combine the merits of both. 
	
	It is worth noting that optimization in fading channels can lead to many interesting research problems. 
	For instance, current LDIP algorithm randomly picks up a vertex to be the R-Vertex when there exist multiple candidates of R-Vertices. 
	This is however too sub-optimum in fading channels where the max-min principle should be incorporated in the graph partitioning.  
	Moreover, communication channels can be frequency selective. 
	Therefore, sub-channel selection mechanism should be incorporated to take the advantage of channel frequency diversity-gain.
	Our solution to those research problems will be presented in the journal paper version of this work. 
	
	\section{Experiment Design and Simulation Results}
	\subsection{Model for Generating Robot Subsets}\label{sec4a}
	Computer simulations are used to evaluate the proposed subset selection approach in terms of the lifetime of robot swarm. 
	The major challenge set to computer simulations is forming robot subsets in Monte Carlo trials. 
	This is because neither stochastic subset models nor deterministic models are available in the literature.
	Deterministic models are relatively easier to develop; however they are not generic and representative. 
	This is not a big issue for practical robot swarms since subsets can be formed based on robots' locations.
	More specifically, closely located robots have their data strongly correlated, and they should belong to different subsets. 
	On the other hand, those very distanced robots have their data weakly correlated, and they can together form a subset. 
	The MEC server can form subsets based on robots' location information as well as their empirical data of subset forming. 
	However, this idea is not readily applicable to computer simulations because it requires a meaningful and well verified stochastic-geometry model of robot geo-distribution and location-related data correlations. 
	
	As the early root of the multi-robot edge computing research, we propose a novel and simple stochastic method to generate the robot subsets. 
	Our aim is to divide $N$ robots into $M$ subsets, with each having at least $K$ robots. 
	The inequality ($N\leq MK$) must hold as it is a necessary condition for not excluding any robot in subsets forming. 
	Our subset forming algorithm includes the following three steps:
	
	{\bf\em Step 1 (Initialization):} 
	Set $K$ bins; each bin is given one robot which is randomly drawn from $N$ robots with an equal probability.  
	
	{\bf\em Step 2 (Forming observation groups):} 
	For the leftover $(N-K)$ robots, we randomly throw them into $K$ bins with an equal probability. 
	It is assumed that robots in the same bin have very correlated data (i.e., correlated observations);
	any $K$ robots that are drawn from different bins carry sufficient information for the edge computing; as specified in {\em Definition \ref{def2}}.
	
	{\bf\em Step 3 (Forming robot subsets):} 
	Denote $M_k$ the number of robots in the $k^{th}$ bin. 
	There are three possible cases: {\em a)} $M_k=M$; {\em b)} $M_k>M$; {\em c)} $M_k<M$.
	
	There is no problem with the case {\em a)} since all $M_k$ robots can be evenly allocated to $M$ robot subsets.
	For the case {\em b)}, each subset should take at least $J_k\triangleq\lfloor(M_k)/(M)\rfloor$ robots from the $k^{th}$ bin, 
	and then we randomly choose $(M_k-J_kM)$ subsets and allocate one more robot to each of them. 
	For the case {\em c)}, we randomly duplicate robots in the $k^{th}$ bin.
	For instance, there is a bin accommodating two robots labelled by $\{1, 2\}$.
	Assuming $M=4$, we can randomly generate a set $\{1, 2, 1, 1\}$, $\{1, 2, 2, 2\}$, or $\{1, 2, 1, 2\}$ for simulation uses (i.e., making the case $M_k=M$).
	
	This stochastic model can cover almost all possible cases of robot subset forming, including those illustrated in \figref{fig2}.
	
	\subsection{Parameter Setting for Simulations}
	In addition to the forming of robot subsets, there are many other parameters requiring appropriate setting.
	Those parameters include: $\mathcal{E}_n, \varepsilon, \epsilon, P_n$. 
	One of challenging issues is to determine the relationship between the energy costs for signal transmission, for modulation and encoding, and for robot task execution, respectively. 
	This would give us a unified energy consumption model for computer simulations.
	Unfortunately, we are not aware of any explicit description of their relationships in the literature despite a very extensive survey. 
	Therefore, in our simulations, we assume that a robot's energy is divided into two fixed and separated parts, one for communication and the other for task execution. 
	In order to focus our study in the communication domain, it is assumed that the bottleneck of robot swarm is the communication energy. 
	In other words, when a robot has its communication energy used up, this robot counts as `dead'.
	
	To simplify the computer simulations, we set $\mathcal{E}_n$ and $P_n$ to be identical for all robots and omit the subscript $(\cdot)_n$.
	Moreover, we ignore the energy cost for modulation and coding (i.e., $\epsilon$) since they are often negligibly small when comparing to the radio transmission energy. 
	The term $\mathcal{E}$ is now only referred to the total energy for communications. 
	In our experiments, it is set to allow a robot conducting $200$ data transmissions in the AWGN channel with the SNR of $10$ dB.
	This essentially sets a robot's lifetime in the AWGN channel (i.e., $i=200$). 
	
	\subsection{Simulation Results and Discussion}
	Our computer simulations mainly include two experiments, concerning communication channels to be AWGN and independent and identically distributed (i.i.d.) Rayleigh, respectively. 
	We consider two baselines: 
	\begin{itemize}
		\item Conventional single-user equivalent edge computing, where all robots send their data to the MEC server \cite{8279411}. 
		In the AWGN channel, this baseline approach straightforwardly gives the swarm lifetime of $i=200$ according to our setting, and we aim to demonstrate the gain of exploiting source correlation in the multi-user edge computing.
		\item Max-min algorithm for subset selection. 
		This algorithm is extended from the max-min user selection approach in sensor networks \cite{8955967}, but we apply it on the subset level.
		Strictly speaking, this is a novel algorithm, and we use it as a baseline to demonstrate advantages of the proposed LDIP algorithm. 
	\end{itemize}
	The approach under evaluation is the LDIP-enabled subset selection algorithm (see  \eqref{eqn32} for the generic form).
	The one with only R-Vertices involved in the subset selection is named LDIP-R-Vertices,
	and the one with all vertices involved in the subset selection is named LDIP-All.
	
	\subsubsection*{Experiment 1}
\begin{figure}[t]
		\centering
		\includegraphics[scale=0.32]{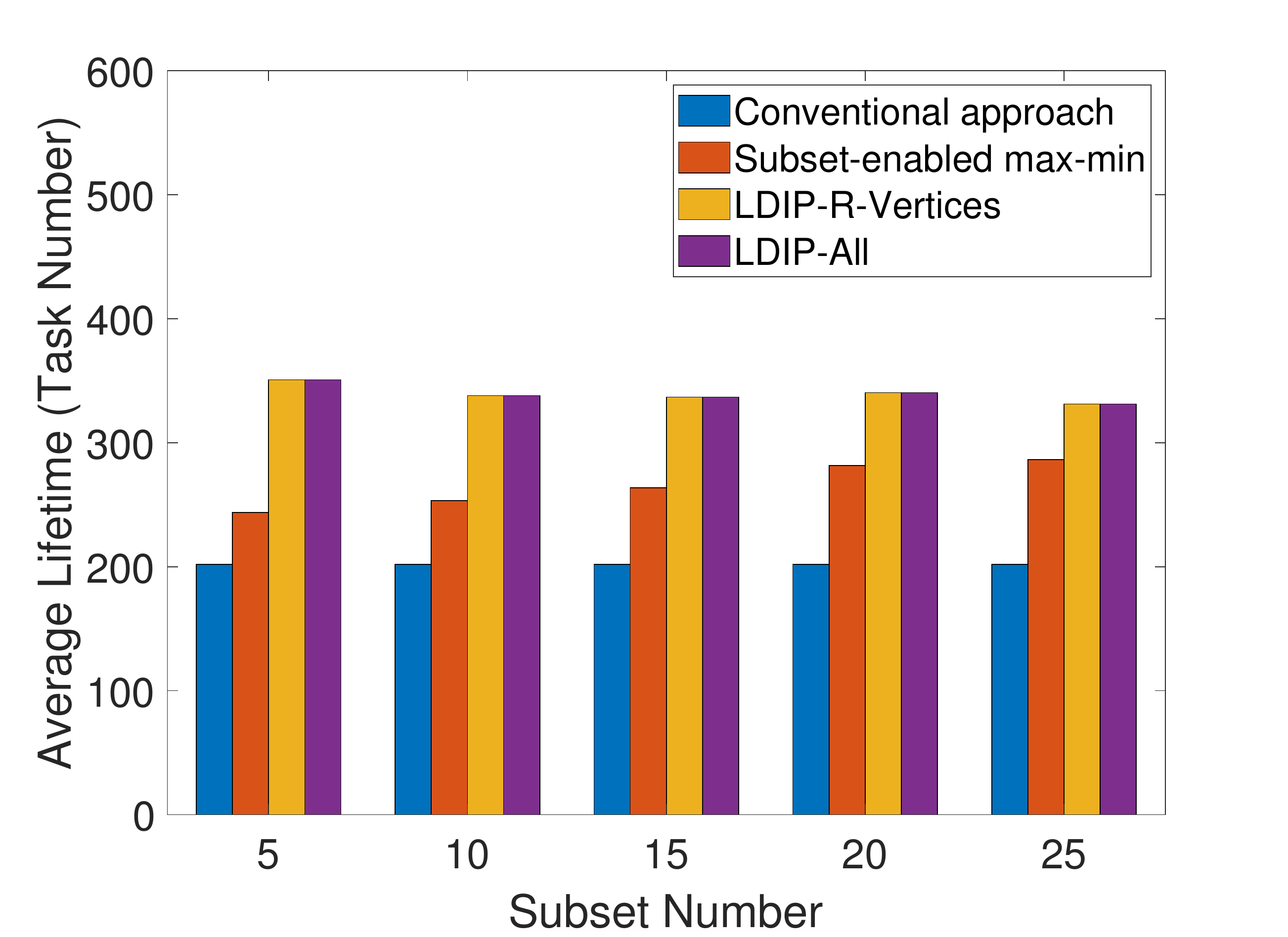}
		\vspace{-0.4em}
		\caption{The average swarm lifetime as a function of $M$ in AWGN channels.}
		\label{fig3}
		\vspace{-1.3em}
	\end{figure}

	The objective of this experiment is to evaluate the proposed approach in the AWGN channel. 
	We simulate a robot swarm consisting of $(N=30)$ robots.
	Each robot subset has at least ($K=8$) robots.
	\figref{fig3} illustrates the swarm lifetime (averaged over $500$ Monte Carlo trials) as a function of the number of subsets ($M$).
	It is shown that subset-selection approaches can improve the swarm lifetime by $25\%\sim75\%$.
	Such a significant gain is due to the exploitation of data source correlation. 
	
	It can also be observed that LDIP approaches outperform the max-min approach in the subset selection. 
	This is because, in the AWGN channel, \eqref{eqn32} reduces to \eqref{eqn17}, where the round-robin subset selection strategy is optimum (see {\em Theorem \ref{thm01}}), and the max-min principle does not offer any additional gain.
	For the same reason, LDIP-R-Vertices and LDIP-All show identical performances. 
	
	\subsubsection*{Experiment 2}
\begin{figure}[t]
		\centering
		\includegraphics[scale=0.32]{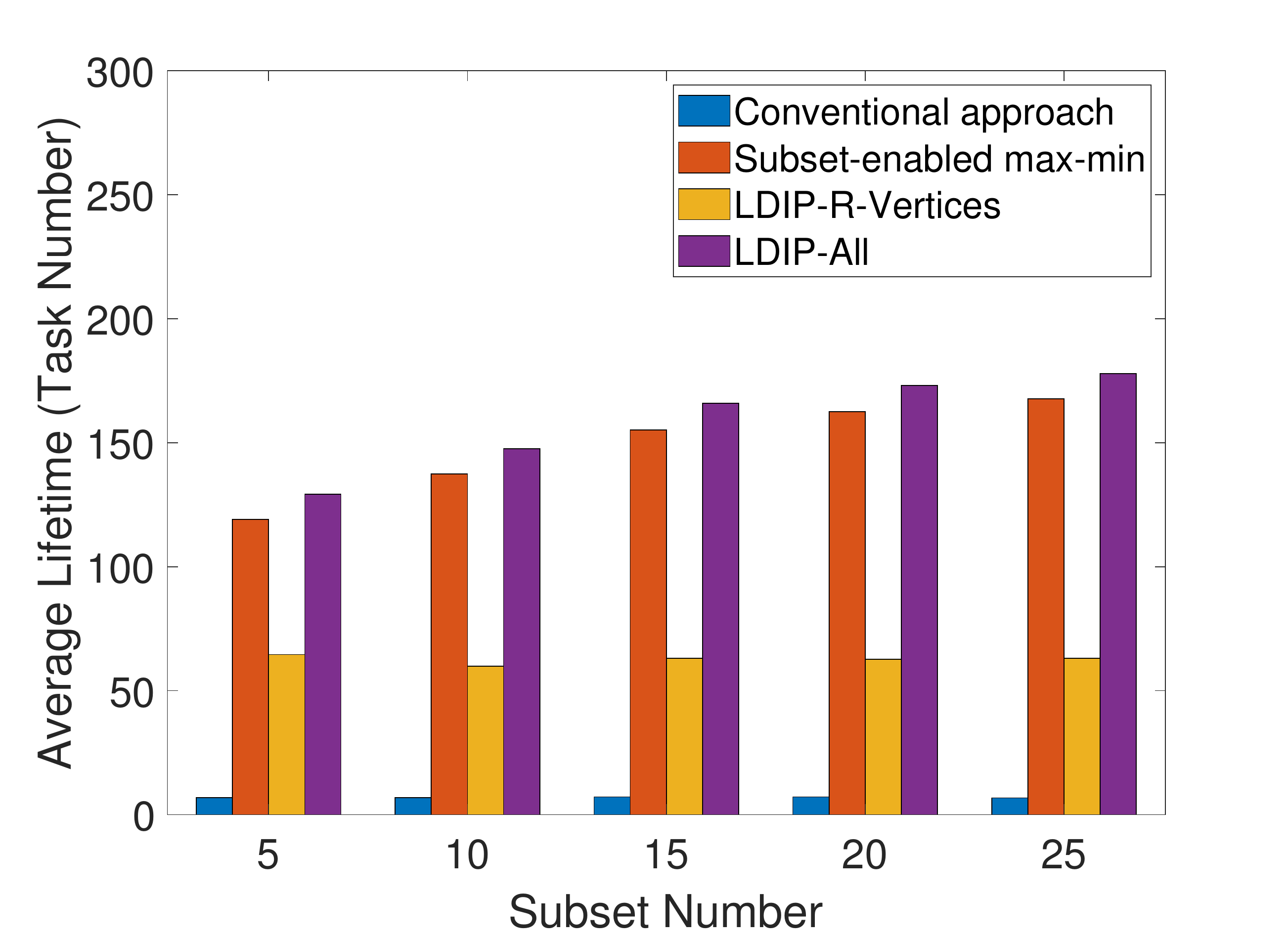}
		\vspace{-0.4em}
		\caption{The average swarm lifetime as a function of $M$ in Rayleigh channels.}
		\label{fig4}
		\vspace{-1.3em}
	\end{figure}

	With the same system setup as in {\em Experiment 1}, this experiment is interested in the performance in  i.i.d. Rayleigh channels.
	{\em Experiment 2} is different from {\em Experiment 1} mainly in two folds: {\em 1)} \eqref{eqn32} does not reduce to \eqref{eqn17}, and thus the max-min principle plays a vital role; 
	{\em 2)} power adaptation is needed in fading channels (see {\em Criterion 1}).
	In our simulations, the transmit power ($p_n$) has a power cap, which is set three times of the transmit power in the AWGN channel. 
	The simulation results are plotted in \figref{fig4}.
	It is observed that the conventional approach has the swarm lifetime significantly reduced. 
	This is because robots have to pay much more transmit power to combat channel fades. 
	Compared to the results in AWGN channels, subset-selection approaches also have the swarm lifetime largely reduced. 
	However, their performances are significantly better than the conventional approach. 
	Another remarkable phenomenon is that LDIP-All offers the best performance, and LDIP-R-Vertices shows worse performance than the max-min approach.
	This implies that the max-min principle is vital in fading channels, and all subsets must be considered in the subset selection.
	LDIP-All combines the merits of LDIP and max-min principles, and thus achieves the best performance. 
	
	\section{Conclusion and Outlook}
	In this paper, a novel subset selection concept has been presented for multi-user edge computing, 
	with the aim to maximize the lifetime of robot swarm through efficient exploitation of the data source correlation. 
	Major contribution of this work includes a subset model describing the data source correlation between robots, an objective function for lifetime optimization, as well as a graph partitioning-based optimization approach. 
	All of the proposed models and approaches are novel, and they have demonstrated remarkable gains both in the AWGN and Rayleigh fading channels. 
	
	Moreover, we have highlighted the future direction towards this topic, which includes the LDIP optimization and user scheduling in frequency-selective fading channels. 
	
	%\balance

	\bibliographystyle{IEEEtran}	
	\bibliography{ref.bib}
\end{document}